\documentclass{PoS}
\usepackage{amsmath}
\usepackage{amssymb}

\newcommand{\bi}  {\begin{itemize}}
\newcommand{\ei}  {\end{itemize}}
\newcommand{\be}  {\begin{enumerate}}
\newcommand{\ee}  {\end{enumerate}}

\newcommand{\bc}  {\begin{center}}
\newcommand{\ec}  {\end{center}}

\newcommand{\beq}{\begin{equation}}
\newcommand{\eeq}{\end{equation}}
\newcommand{\bea}{\begin{eqnarray}}
\newcommand{\eea}{\end{eqnarray}}

\newcommand {\tr}{\mbox{Tr}}

\title{Electric charge susceptibility in 2+1 flavour QCD on an 
anisotropic lattice}

\ShortTitle{Electric charge susceptibility in 2+1 flavour QCD on an 
anisotropic lattice}

\author{\speaker{Pietro Giudice}\\
  Universit\"at M\"unster, Institut f\"ur Theoretische Physik,
  M\"unster, Germany\\
  E-mail: \email{p.giudice@uni-muenster.de}}

\author{Gert Aarts\\
  Department of Physics, College of Science, 
  Swansea University, Swansea, United Kingdom\\
  E-mail: \email{g.aarts@swan.ac.uk}}

\author{Chris Allton\\
  Department of Physics, College of Science, 
  Swansea University, Swansea, United Kingdom\\
  E-mail: \email{c.allton@swan.ac.uk}}

\author{Alessandro Amato\\
  Department of Physics, College of Science, 
  Swansea University, Swansea, United Kingdom\\
  Institute for Theoretical Physics, Universit\"at Regensburg, 
  D-93040 Regensburg, Germany\\
  E-mail: \email{Alessandro.Amato@physik.uni-regensburg.de}}

\author{Simon Hands\\
  Department of Physics, College of Science, 
  Swansea University, Swansea, United Kingdom\\
  E-mail: \email{s.hands@swansea.ac.uk}}

\author{Jon-Ivar Skullerud\\
  Department of Mathematical Physics, National University of Ireland Maynooth, 
  Maynooth, County Kildare, Ireland\\
  E-mail: \email{jonivar@thphys.nuim.ie}}

\abstract{
The FASTSUM Collaboration presents its first results for the electric 
charge susceptibility in QCD 
using $2+1$ dynamical flavours of Wilson quark on anisotropic lattices. 
Spatial volumes of $(2.94 \ \mbox{fm})^3$ are used at fixed cut-off with 
temperatures ranging from below $T_c$ to $\sim 2 T_c$.
}

\FullConference{31st International Symposium on Lattice Field Theory 
  LATTICE 2013\\
  July 29 - August 3, 2013\\
  Mainz, Germany}

\begin{document}

\section{Introduction}

The measurement of fluctuations of conserved charges can probe 
both the thermal state of the medium and its critical behaviour.
The fluctuations are quantified by the susceptibilities, defined
as second (and higher) derivatives of the free energy with respect
to the chemical potential associated with the investigated charge.
In QCD, taking into account  the three light flavours, usually three charges 
are studied in the literature: baryon number, electric charge, strangeness. 
Their susceptibilities probe the actual degrees of freedom that carry 
such charges, {\it i.e.} quarks or their bound states.

QCD shows a very interesting phase structure~\cite{Fukushima:2010bq}.
The main characteristic, at zero baryon chemical potential ($\mu_B$), is a 
crossover transition from a confined, chirally broken phase at low 
temperature ($T$) to a deconfined, chirally symmetric phase at high temperature 
called quark-gluon plasma (QGP) around $T_c \approx 150$MeV.
Fluctuations can be used to probe quark deconfinement~\cite{Asakawa:2000wh}
by studying event-by-event fluctuations of charged 
particle ratios~\cite{Bleicher:2000ek}.
Susceptibilities show a rapid rise in the crossover region: at low 
temperature they are small since quarks are confined; at high
temperature they are large and they approach the ideal gas limit.

There are many specific reasons why physicists are interested in 
studying these quantities; here we recall the following:
\begin{itemize}
\item the QCD transition is today reproduced in the heavy ion collision 
experiments at BNL (RHIC) and at CERN (LHC) where the fluctuations
can be studied~\cite{Abelev:2012pv}: 
it is clearly very important for our understanding of the strong interaction 
to compare their measurements with lattice QCD determinations;

\item susceptibilities can be seen as coefficients of the Taylor expansion
of the free energy with respect to the chemical potentials; in this form
they can be used to perform numerical studies at finite density 
QCD~\cite{Allton:2005gk, Miao:2008sz};

\item QCD may have a critical point, an endpoint of a line of first order phase 
transitions which extends from the $\mu_B$-axis into the ($\mu_B$,$T$) plane.
Susceptibilities can probe its presence and, in particular, the
baryon number susceptibility is expected to diverge at the critical point.
Its position can be estimated studying the radius of 
convergence~\cite{Gavai:2008zr}
of the Taylor series introduced in the previous paragraph;

\item the susceptibilities of the conserved charges enter the
study of the transport properties of the QGP via Kubo 
formulae~\cite{Meyer:2011gj}, {\it  e.g.} the electric charge susceptibility
$\chi_Q$ connects the charge diffusion coefficient $D$ with the electrical 
conductivity $\sigma$: $\sigma=\chi_Q D$.
\end{itemize}
Among the available theoretical tools, lattice calculations are probably 
the best for calculating the properties of QGP and in particular the 
susceptibilities from first principles: this approach is followed
in this work.
Nearly all lattice studies of susceptibilities have so far
been carried out using staggered fermions. In this study we employ instead
clover-improved Wilson fermions (for an earlier study using Wilson
fermions see Ref.~\cite{Borsanyi:2012uq}).

\section{Simulation details}

The action  that we have used is, for the gauge sector, Symanzik-improved 
with tree-level tadpole-improved coefficients and for the fermion sector, 
a $2+1$ flavour anisotropic clover action with stout-link smearing; 
for details see Ref.~\cite{Edwards:2008ja}.
The configurations have been generated using the Chroma 
Library~\cite{Edwards:2004sx} on IBM Blue Gene/P and Blue Gene/Q
supercomputers.

The volumes we have used are $24^3 \times N_t$, where the values of $N_t$,
together with the corresponding temperature and the number of configurations
generated, are given in Table~\ref{table1}.
\begin{table}[htbp]
\centering
\vspace{4mm}
  \begin{tabular}{|l|c|c|c|c|c|c|c|}
    \hline
    $N_t$    & 40  & 36  & 32   & 28   & 24   & 20    &   16 \\
    \hline
    $T$ [MeV]& 141 & 156 & 176 & 201 & 235 & 281 & 352 \\
     \hline
    $N_{\text{cfg}}$ & 500 & 500 & 1000 & 1000 & 1000 &  1000 & 1000 \\ 
    \hline
  \end{tabular}
  \caption{Some simulation parameters.}
    \label{table1}
\vspace{2mm}
\end{table} 

We used anisotropic lattices with a physical anisotropy 
$\xi \equiv a_s/a_t=3.5$: 
the temporal lattice spacing, measured using $M_\Omega$ to set the scale,  
is $a_t = 0.03506(23)$~fm ($a_t^{-1}=5.63(4)$~GeV) and $a_s = 0.1227(8)$~fm;
for details see Ref.~\cite{Lin:2008pr}. 
Moreover, $M_\pi/M_\rho=0.446(3)$ and $M_\pi = 392(4)$~MeV, {\it i.e.} 
2.9 times bigger than the physical value.

\section{Definitions and observables}

Introducing the quark chemical potentials $\mu_u, \mu_d, \mu_s$ and the charges
$q^u=+\frac{2}{3}$, $q^d=q^s=-\frac{1}{3}$, for each
flavour (we use also here the correspondence:  $1\leftrightarrow$up, 
$2\leftrightarrow$down, $3\leftrightarrow$strange and also in the following
$l\rightarrow$ light, $s\rightarrow$ strange) we define the 
following quantities:
\begin{itemize}
\item quark number densities: $n_i=\frac{T}{V} 
\frac{\partial \ln{Z}}{\partial \mu_i}$;
\item quark number susceptibilities: $\chi_{ij}=\frac{T}{V} 
\frac{\partial^2 \ln{Z}}{\partial \mu_i\partial \mu_j }$.
\end{itemize}
Introducing the electric charge chemical potential $\mu_Q$, the 
electric charge $Q$ and its susceptibility $\chi_Q$ are given by:
\beq
Q=\frac{T}{V} \frac{\partial \ln{Z}}{\partial \mu_Q}=\sum_{i=1}^3  q^i n_i \ ,
\quad\quad
 \chi_Q=\frac{\partial Q}{\partial \mu_Q}=
\sum_{i=1}^3 (q^i)^2  \chi_{ii} + \sum_{i\ne j}^3 q^i q^j \chi_{ij} .
\eeq
%
We now introduce the following terms ($M$ is the fermion matrix):
\beq
T_1^i=
\langle \frac{T}{V} \tr{\left[ M^{-1} \frac{\partial M}{\partial \mu_i} 
\right]}\rangle \ ,
\quad
T_2^i=
\langle \frac{T}{V} \tr{\left[ M^{-1} \frac{\partial^2 M}{\partial \mu_i^2} 
\right]} \rangle \ , \nonumber
\eeq
\beq
T_3^{ij}=
\langle \frac{T}{V} \tr{\left[ M^{-1} \frac{\partial M}{\partial \mu_i} 
\right]}  \tr{\left[ M^{-1} \frac{\partial M}{\partial \mu_j} 
\right]} \rangle \ ,
\quad
T_4^i=
\langle \frac{T}{V} \tr{\left[ M^{-1} \frac{\partial M}{\partial \mu_i} 
M^{-1} \frac{\partial M}{\partial \mu_i} \right]}\rangle \ ,
\label{terms}
\eeq
which we can be used to determine the quantities: 
\bea
n_i&=&T_1^i \ ; \\
\chi_{ii}&=&-(T_1^i)^2+T_2^i+T_3^{ii}-T_4^i \ ;  \\
\chi_{ij}&=&-T_1^i T_1^j+T_3^{ij} \quad \mbox{(here $i \ne j$)} \ .
\eea
Note that $T_1^i=0$ for zero chemical potentials, in fact $n_i=0$ 
and $Q=0$; therefore there is no need to compute this quantity
numerically.

The simplest quantity we can determine is the isospin susceptibility
(defining $\mu_I=\mu_d-\mu_u$):
\beq
\chi_I=\frac{T}{V}\frac{\partial^2 \ln{Z}}{\partial \mu_I^2}=
\frac{1}{4} 
\frac{T}{V} \left[ \frac{\partial^2 \ln{Z}}{\partial \mu_u^2}+
\frac{\partial^2 \ln{Z}}{\partial \mu_d^2}-2 
\frac{\partial^2 \ln{Z}}{\partial \mu_u \partial \mu_d }
\right]=\frac{1}{4} \sum_{i=1}^2 \left[ T_2^i-T_4^i \right] .
\eeq
We note that it depends only on the terms $T_2$ and $T_4$
but not $T_3$ which, from a numerical point of view, is the most 
expensive quantity to determine since it comes from a disconnected 
diagram.

Below we compare our results at high temperature with
the expressions for free massless fermions in the continuum limit, which
read:
\bea
\chi_I &\rightarrow& \frac{T^2}{2} ; \label{eq:chiI} \\
\chi_Q &\rightarrow& T^2 \sum_{i=1}^3 q_i^2 \ ,  \label{eq:chiQ}
\eea
where $\sum_{i=1}^3 q_i^2=2/3$. 

Moreover, we also compare the susceptibilities with those obtained from 
free Wilson fermions, which can be determined analytically starting 
from the following expression for the baryon density, valid for $N_f$
flavours and $N_c$ colours:
\beq
n^{free}_B(\mu)=\frac{8 N_f N_c}{N_s^3 N_t} \sum_{k_4, k_i} 
\frac{i P_4 \sin{(k_4+i\mu)} 
\left[ 1- 2 P_s \sum_i \cos{k_i} \right] }
{ \left[ 
1 -  2 P_s \sum_i \cos{k_i}- 2 P_4 \cos{(k_4+i\mu)} \right]^2 + 
4 P_s^2 \sum_i \sin^2{k_i}+ 4 P_4^2 \sin^2{(k_4+i\mu)}} \ ,\nonumber
\label{nlattwilson}
\eeq
where $i=1,2,3$, $ \ k_i$ are the discretised momenta, the coefficients
$P_4$ and $P_s$ are given by: \\
\beq
P_4= \nu_t/(m+\nu_t+3/\gamma_f )\ ,  \quad
P_s= \nu_s/ \left[ 2 \ \xi_0 \ (m+\nu_t+3 / \gamma_f ) \right] , \nonumber
\eeq
and $\nu_t=1$, $\nu_s=\gamma_g/\gamma_f$ and $\xi_0=\gamma_g$ 
($\gamma_g$ is the bare gauge anisotropy and $\gamma_f$ is the bare fermion 
anisotropy). Note that when we compare with our data, $\gamma_f$ has
to be replaced with $\xi$, the physical value.
This relation was determined following the approach discussed
in Sec.~4.2.4 of Ref.~\cite{Montvay:1994cy}.

\section{A few technical aspects}

We estimated the traces of Eq.~(\ref{terms}) using $N_v$ noise vectors $\eta_i$;
for connected terms, {\it i.e.} $T_2, T_4$, we used just $N_v=9$ noise 
vectors but 
for the disconnected term, {\it i.e.} $T_3$, we used $N_v=200$ noise vectors 
for $N_t=40$ and $N_v=100$ for the others. As discussed in 
Ref.~\cite{Gottlieb:1987ac} the term $T_3$,
which contain two traces, has a significantly larger variance than the other
terms and therefore needs a larger number of random vectors. 
It is worth noting that in this case the standard error of the mean 
falls like the inverse of the number of noise vectors (not like the square 
root of it) so that increasing $N_v$ has an evident effect in the final result.

As discussed in Ref.~\cite{Gavai:2001ie} the most efficient way
to determine the square of a trace is by the following relation:
\beq
(\tr{A})^2=\frac{2}{N_v(N_v-1)}\sum_{i>j=1}^{N_v} \eta^\dagger_i A \eta_i \
\eta^\dagger_j A \eta_j \ ,
\eeq
{\it i.e.} the diagonal terms are not taken into account because they would
introduce a bias in the final result given by a term whose relative 
significance is $O(1/N_v)$. 

We tested the effect of \emph{dilution}~\cite{Wilcox:1999ab, Foley:2005ac}
in the colour and Dirac space but we did not see any improvement in the final
results. A dilution test done with $N_t=40$ has brought the following results:
we have noted a reduction of the errors in the real part of $T_1$ by
a factor $\sim 2$, and in the imaginary part of $T_2$ by a factor $\sim 3$; 
in all other cases the effect of dilution has been counterproductive.
This is not a surprise because dilution has a positive effect only if 
the off-diagonal elements of a matrix dominate the diagonal ones, 
therefore the effect depends strongly on the observable under 
consideration\footnote{See Eq.~(20) of Ref.~\cite{Wilcox:1999ab}.}.

\section{Results}

\begin{figure}[ht]
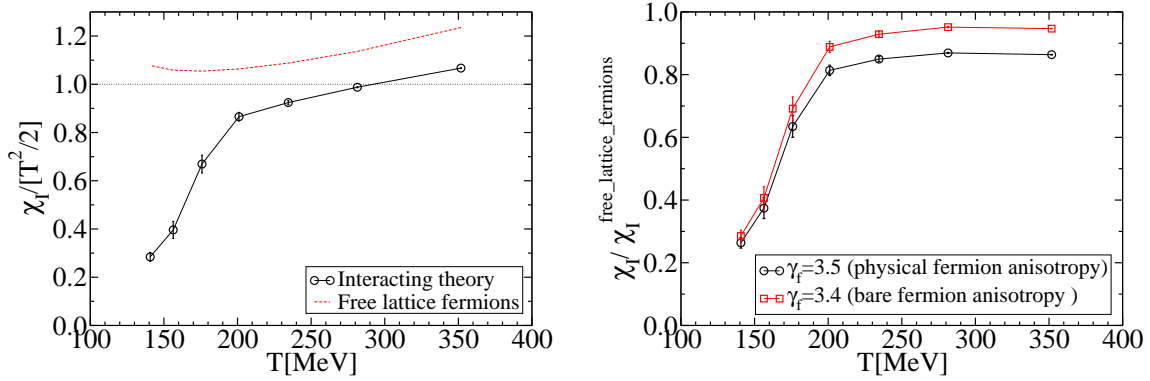

  \begin{center}
    \vspace{4mm}
    \includegraphics[width=0.48\hsize,angle=0]{./plots/chiISOSPIN_vs_highT.eps}
    \hspace{0.4cm}
    \includegraphics[width=0.48\hsize,angle=0]{./plots/chiISOSPIN_vs_free.eps}
    \caption{(Left) $\chi_I$ normalised using the high temperature
expected behaviour (for free massless fermions living in a continuum 
spacetime). For comparison the same quantity for free massless lattice 
fermions is plotted. 
(Right) $\chi_I$ normalised with respect to the same quantity calculated for 
free massless lattice fermions using two values for the fermion anisotropy.}
    \label{plot1}
  \end{center}
\end{figure}

In this Section we present our preliminary results. 
In Figure~\ref{plot1} (Left) we present the result for the isospin 
susceptibility, normalising it with respect to the continuum expectation 
at high temperature, Eq.~(\ref{eq:chiI}). 
As can be seen for $T \gtrsim 270$MeV the value is systematically higher than 
the expected one, {\it i.e.}  unity, for free massless fermions living 
in a continuum spacetime.
This is a lattice artefact which is a consequence of combined effect 
of our fixed cut-off approach and the small 
number of \emph{effective} sites in the temporal direction 
($N_t/\xi=4.6, 5.7, 6.8, \ldots$). This is evident looking at the dashed-red 
line which shows the same quantity for free massless lattice fermions, 
{\it i.e.} taking into account the effect of discretisation of the spacetime.

In Figure~\ref{plot1} (Right) we compare directly the result with that of 
massless free Wilson fermions. Here the discretisation effects are clearly 
taken into account and for  $T \gtrsim 250$MeV the value of $\chi_I$ is 
above 85\% of the Stefan-Boltzmann value. In the figure we plot moreover
the result with two values of the fermion anisotropy: 
the physical one, $\gamma_f=\xi=3.5$, and the bare one, $\gamma_f=3.4$.
The importance of using the correct value for the parameter
$\gamma_f$ to compare our results with the Stefan-Boltzmann limit is evident.

\begin{figure}[ht]
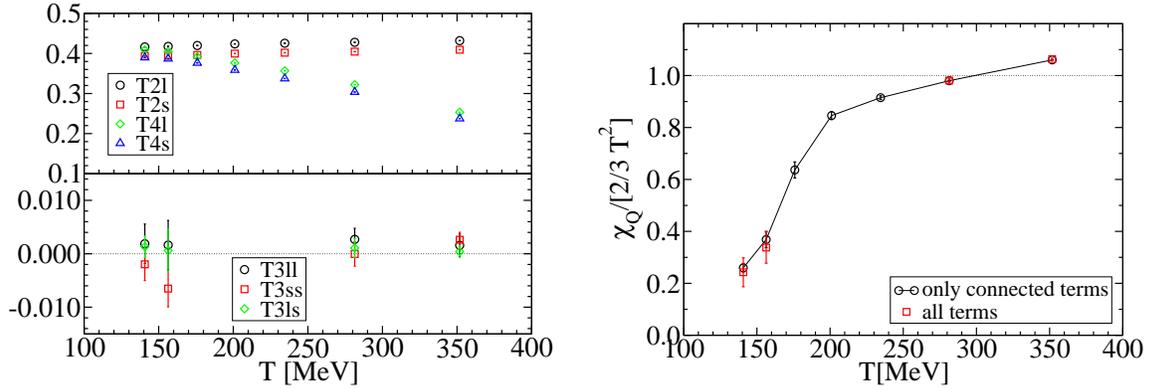

  \begin{center}
    \vspace{4mm}
    \includegraphics[width=0.48\hsize,angle=0]{./plots/terms.eps}
    \hspace{0.4cm}
    \includegraphics[width=0.48\hsize,angle=0]{./plots/chi_electric_conn.eps}
    \caption{(Left) Terms $T_2$, $T_4$ and $T_3$. (Right) $\chi_Q$ 
normalised using the expected high temperature behaviour (for free massless 
fermions living in a continuum spacetime).}
    \label{plot2}
  \end{center}
\end{figure}

The contribution of the different terms to the susceptibilities
are shown in Figure~\ref{plot2} (Left). We can see that the main
contribution to the final value of the susceptibilities is given
by the terms $T_2$ and $T_4$; note also the appreciable difference
between the light and the strange quarks.

From Figure~\ref{plot2} (Left) we can see moreover the contribution of the 
disconnected term $T_3$. As discussed in Ref.~\cite{Gottlieb:1987ac}
it is expected that this term is negative, basically as a consequence 
of hopping parameter expansion analysis.
Hard thermal loop (HTL) perturbation theory showed a decade 
ago~\cite{Blaizot:2001vr} that $\chi_{ij}=T_3^{ij}$ should be different 
from zero, showing a clear correlation between different flavours.
Recent lattice calculations~\cite{Borsanyi:2011sw} 
have shown a clear dip for the off-diagonal term (in the crossover region). 
Our results, both for diagonal and off-diagonal contributions, show that 
it is compatible with zero, either at low and high temperature: 
this is probably a consequence of our relatively large pion mass.

The electric charge susceptibility is presented in 
Figure~\ref{plot2} (Right). Here we present this quantity normalised
with respect to the high temperature behaviour expected for free massless 
fermions living in a continuum spacetime, Eq.~(\ref{eq:chiQ}). 
The cut-off effects are
clearly present at high temperature as for Figure~\ref{plot1} (Left).
The rapid rise of $\chi_Q$, signalling the transition from the confined 
phase to the deconfined phase, is evident around $T_c \sim 150$MeV.

In this preliminary figure,  we have included all terms of Eq.~(\ref{terms}) 
only for a few values of the temperature (see legend).
Taking into account the results of Figure~\ref{plot2} (Left)
and the fact that the disconnected contribution has had little effect 
at those temperatures where the electric charge susceptibility has been 
calculated, we can infer it will also have little affect for the other 
temperatures.

\section{Conclusion and outlook}

We have presented here our preliminary calculations of isospin and
electric charge susceptibility in $2+1$ flavour QCD on 
anisotropic lattices using a fixed cut-off approach to explore a 
range of temperature which goes from below $T_c$ to $\sim 2 T_c$.
We are completing the determination of the susceptibilities on the volume 
$24^3$ on all the configurations at our disposal.
When this is complete, we plan to combine these results
with the measurement of the electrical conductivity, see 
Refs.~\cite{Amato:2013naa},
to determine the charge diffusion coefficient $D$.
Moreover, by an appropriate combination of the terms of Eq.~(\ref{terms}),
we will determine the other relevant susceptibilities, in particular the one
associated with the baryon number, $\chi_B$.


\begin{thebibliography}{99}

\bibitem{Fukushima:2010bq}
  K.~Fukushima and T.~Hatsuda,
  Rept.\ Prog.\ Phys.\  {\bf 74} (2011) 014001
  [arXiv:1005.4814 [hep-ph]].


\bibitem{Asakawa:2000wh}
  M.~Asakawa, U.~W.~Heinz and B.~Muller,
  Phys.\ Rev.\ Lett.\  {\bf 85} (2000) 2072
  [hep-ph/0003169].


\bibitem{Bleicher:2000ek}
  M.~Bleicher, S.~Jeon and V.~Koch,
  Phys.\ Rev.\ C {\bf 62} (2000) 061902
  [hep-ph/0006201].


\bibitem{Abelev:2012pv}
  B.~Abelev {\it et al.}  [ALICE Collaboration],
  Phys.\ Rev.\ Lett.\  {\bf 110} (2013) 152301
  [arXiv:1207.6068 [nucl-ex]].


\bibitem{Allton:2005gk}
  C.~R.~Allton, M.~Doring, S.~Ejiri, S.~J.~Hands, O.~Kaczmarek, F.~Karsch, E.~Laermann and K.~Redlich,
  Phys.\ Rev.\ D {\bf 71} (2005) 054508
  [hep-lat/0501030].


\bibitem{Miao:2008sz}
  C.~Miao {\it et al.}  [RBC-Bielefeld Collaboration],
  PoS LATTICE {\bf 2008} (2008) 172
  [arXiv:0810.0375 [hep-lat]].


\bibitem{Gavai:2008zr}
  R.~V.~Gavai and S.~Gupta,
  Phys.\ Rev.\ D {\bf 78} (2008) 114503
  [arXiv:0806.2233 [hep-lat]].


\bibitem{Meyer:2011gj}
  H.~B.~Meyer,
  Eur.\ Phys.\ J.\ A {\bf 47} (2011) 86
  [arXiv:1104.3708 [hep-lat]].


\bibitem{Borsanyi:2012uq}
  S.~Borsanyi, S.~Durr, Z.~Fodor, C.~Hoelbling, S.~D.~Katz, S.~Krieg, D.~Nogradi and K.~K.~Szabo {\it et al.},
  JHEP {\bf 1208} (2012) 126
  [arXiv:1205.0440 [hep-lat]].


\bibitem{Edwards:2008ja}
  R.~G.~Edwards, B.~Joo and H.~-W.~Lin,
  Phys.\ Rev.\ D {\bf 78} (2008) 054501
  [arXiv:0803.3960 [hep-lat]].


\bibitem{Edwards:2004sx}
  R.~G.~Edwards {\it et al.}  [SciDAC and LHPC and UKQCD Collaborations],
  Nucl.\ Phys.\ Proc.\ Suppl.\  {\bf 140} (2005) 832
  [hep-lat/0409003].


\bibitem{Lin:2008pr}
  H.~-W.~Lin {\it et al.}  [Hadron Spectrum Collaboration],
  Phys.\ Rev.\ D {\bf 79} (2009) 034502
  [arXiv:0810.3588 [hep-lat]].


\bibitem{Montvay:1994cy}
  I.~Montvay and G.~Munster,
  Cambridge, UK: Univ. Pr. (1994) 491 p. (Cambridge monographs on 
mathematical physics).


\bibitem{Gottlieb:1987ac}
  S.~A.~Gottlieb, W.~Liu, D.~Toussaint, R.~L.~Renken and R.~L.~Sugar,
  Phys.\ Rev.\ Lett.\  {\bf 59} (1987) 2247;
  Phys.\ Rev.\ D {\bf 38} (1988) 2888.


\bibitem{Gavai:2001ie}
  R.~V.~Gavai, S.~Gupta and P.~Majumdar,
  Phys.\ Rev.\ D {\bf 65} (2002) 054506
  [hep-lat/0110032].


\bibitem{Wilcox:1999ab}
  W.~Wilcox,
  hep-lat/9911013.


\bibitem{Foley:2005ac}
  J.~Foley, K.~Jimmy Juge, A.~O'Cais, M.~Peardon, S.~M.~Ryan 
and J.~-I.~Skullerud,
  Comput.\ Phys.\ Commun.\  {\bf 172} (2005) 145
  [hep-lat/0505023].


\bibitem{Borsanyi:2011sw}
  S.~Borsanyi, Z.~Fodor, S.~D.~Katz, S.~Krieg, C.~Ratti and K.~Szabo,
  JHEP {\bf 1201} (2012) 138
  [arXiv:1112.4416 [hep-lat]].


\bibitem{Blaizot:2001vr}
  J.~P.~Blaizot, E.~Iancu and A.~Rebhan,
  Phys.\ Lett.\ B {\bf 523} (2001) 143
  [hep-ph/0110369].


\bibitem{Amato:2013naa}
  A.~Amato, G.~Aarts, C.~Allton, P.~Giudice, S.~Hands and J.~-I.~Skullerud,
  arXiv:1307.6763 [hep-lat]; PoS LATTICE {\bf 2013} (2013) 176.


\end{thebibliography}
\end{document}